# Efficient generation of near infra-red single photons from the zero-phonon line of a single molecule

**J.-B. Trebbia, H. Ruf, Ph. Tamarat, and B. Lounis***

*Centre de Physique Moléculaire Optique et Hertzienne, Université de Bordeaux and CNRS, 351 cours de la Libération, F-33405 Talence (France)*

*[blounis@u-bordeaux1.fr](mailto:blounis@u-bordeaux1.fr)*

**Abstract:** Using the zero-phonon line (ZPL) emission of a single molecule, we realized a triggered source of near-infra-red ($\lambda=785$ nm) single photons at a high repetition rate. A Weierstrass solid immersion lens is used to image single molecules with an optical resolution of 300 nm (~0.4$\lambda$) and a high collection efficiency. Because dephasing of the transition dipole due to phonons vanishes at liquid helium temperatures, our source is attractive for the efficient generation of single indistinguishable photons.

**References and links**

1. B. Lounis, and M. Orrit, "Single-photon sources", Rep. Prog. Phys. **68**, 1129-1179 (2005).
2. Ph. Grangier, B. Sanders, and J. Vuckovic, Editors, "Focus on Single Photons on Demand", New. J. Phys. **6** (2004).
3. C. Brunel, Ph. Tamarat, B. Lounis, and M. Orrit, "Triggered source of single photons based on controlled single molecule fluorescence", Phys. Rev. Lett. **83**, 2722-2725 (1999).
4. B. Lounis, and W. E. Moerner, "Single photons on demand from a single molecule at room temperature", Nature **407**, 491-493 (2000).
5. P. Michler, A. Kiraz, C. Becher, W.V. Schoenfeld, P.M. Petroff, L.D. Zhang, E. Hu, A. Imamoglu, "A quantum dot single-photon turnstile device", Science **290**, 2282-2285 (2000).
6. C. Kurtsiefer, S. Mayer, P. Zarda, and H. Weinfurter, "Stable solid-state source of single photons", Phys. Rev. Lett. **85**, 290-293 (2000).
7. A. Beveratos, S. Kuhn, R. Brouri, T. Gacoin, J.P. Poizat, P. Grangier, "Room temperature stable single-photon source", Eur. Phys. J. D **18**, 191-196 (2002).
8. A. Kuhn, M. Hennrich, and G. Rempe, "Deterministic single-photon source for distributed quantum networking", Phys. Rev. Lett. **89**, 067901 (2002).
9. M. Keller, B. Lange, K. Hayasaka, W. Lange, and H. Walther, "Continuous generation of single photons with controlled waveform in an ion-trap cavity system", Nature **431**, 1075-1078 (2004).
10. C. Santori, D. Fattal, J. Vuckovic, G.S. Solomon, and Y. Yamamoto, "Indistinguishable photons from a single-photon device", Nature **419**, 594-597 (2002).
11. J. Beugnon, M.P.A. Jones, J. Dingjan, B. Darquié, G. Messin, A. Browaeys, and P. Grangier, "Quantum interference between two single photons emitted by independently trapped atoms", Nature **440**, 779-782 (2006).
12. M. Orrit, J. Bernard, R. Brown, and B. Lounis, "Optical spectroscopy of single molecules in solids", Prog. Opt. **35**, 61-144 (1996).
13. T. Basché, W.E. Moerner, M. Orrit and U.P. Wild, Editors, *"Single-Molecule Optical Detection, Imaging and Spectroscopy"*, VCH, Weinheim, Germany, (1997).
14. Ph. Tamarat, A. Maali, B. Lounis, and M. Orrit, "Ten years of single-molecule spectroscopy", J. Phys. Chem. A **104**, 1-16 (2000).
15. F. Kulzer, S. Kummer, R. Matzke, C. Bräuchle, and Th. Basché, "Single-molecule optical switching of terrylene in p-terphenyl", Nature **387**, 688-691 (1997).
16. A. Kiraz, M. Ehrl, T. Hellerer, O.E. Mustecaplioglou, C. Bräuchle, and A. Zumbusch, "Indistinguishable photons from a single molecule", Phys. Rev. Lett. **94**, 223602 (2005).

17. V. Zwiller, H. Blom, P. Jonsson, N. Panev, S. Jeppesen, T. Tsegaye, E. Goobar, M.E. Pistol, L. Samuelson, G. Bjork, "Single quantum dots emit single photons at a time: Antibunching experiments", App. Phys. Lett. **78**, 2476-2478 (2001).
18. R. Lettow, V. Ahtee, R. Pfab, A. Renn, E. Ikonen, S. Götzinger, and V. Sandoghdar, "Realization of two Fourier-limited solid-state single-photon sources", Opt. Express **15**, 15842-15847 (2007).
19. Q. Wu, R.D. Grober, D. Gammon, and D.S. Katzer, "Imaging spectroscopy of two-dimensional excitons in a narrow GaAs/AlGaAs quantum well", Phys. Rev. Lett. **83**, 2652-2655 (1999).
20. V. Zwiller, and G. Bjork, "Improved light extraction from emitters in high refractive index materials using solid immersion lenses", J. Appl. Phys. **92**, 660-665 (2002).
21. K. A. Serrels, E. Ramsay, P. A. Dalgarno, B. D. Gerardot, J. A. O'Connor, R. H. Hadfield, R. J. Warburton, and D. T. Reid, "Solid immersion lens applications for nanophotonic devices," J. Nanophoton. **2**, 021854 (2008).
22. C. Hofmann, A. Nicolet, M.A. Kol'chenko, and M. Orrit, "Towards nanoprobes for conduction in molecular crystals: Dibenzoterrylene in anthracene crystals", Chem. Phys. **318**, 1-6 (2005).
23. A. A. L. Nicolet, P. Bordat, C. Hofmann, M.A. Kol'chenko, B. Kozankiewicz, R. Brown, and M. Orrit, "Single dibenzoterrylene molecules in an anthracene crystal: Main insertion sites", Chem. Phys. Chem. **8**, 1929-1936 (2007).

Over the past decade, significant efforts have been spent in the development of new and reliable triggered single-photon sources [1, 2] based on the controlled emission of single quantum emitters such as organic molecules [3, 4], semiconductor quantum dots [5], color centers in diamond [6, 7], or trapped atoms [8] or ions [9] in the gas phase. Numerous applications for single-photon sources have been envisaged in the field of quantum information. However, an essential element for a photon-based quantum logic toolbox would be a light source that delivers on demand single photons with similar wavepackets. Triggered sources of indistinguishable single photons have been realized using individual quantum dots [10] and individual trapped atoms [11]. Single fluorescent molecules in solids are a promising alternative to these two systems. In comparison to self-assembled quantum dots, the optical coherence lifetime of organic molecules is longer by one order of magnitude. At liquid helium temperatures and for well chosen fluorophore-matrix systems, dephasing of the transition dipole due to phonons vanishes [12]. The purely electronic transition line, also called the zero-phonon line (ZPL), has a spectral width limited by the excited-state lifetime [13]. Such molecules behave like two-level systems [14] with a fluorescence quantum yield close to unity, thus offering optical properties similar to those of trapped single atoms. Moreover, the photostability of organic molecules trapped in crystalline hosts is excellent at cryogenic temperatures, and allows continuous optical measurements over days [15]. It has been shown that the ZPL emission of a cw-excited single molecule produces indistinguishable photons in the visible domain [16].

In this work, we realized a triggered source of single photons in the near infra-red (785 nm) at a high repetition rate (16 MHz), using the strong ZPL emission of single dibenzoterrylene (DBT) molecules embedded in an anthracene (Ac) crystal. The wavelength of the ZPL matches the maximal sensitivity of silicon avalanche photodiodes (APDs) and is suited for propagation in telecommunication fibers. A solid immersion lens (SIL) used in the Weierstrass configuration provides a high collection efficiency of the single photon source, as well as a high spatial resolution microscopy of the single molecules.

To enhance the light extraction efficiency of a point source embedded in a solid-state medium, two configurations based on SILs with index of refraction $n_{SIL}$ matching the refractive index of the host medium have been proposed. Hemispherical SILs have been used to increase the luminescence signal of single self-assembled quantum dots [17]. Since the emitter is at the center of the hemisphere in this geometry, a second optical system (e.g. a lens) with the highest possible numerical aperture (NA) is needed in order to improve the photon collection efficiency and the optical resolution which is given by $0.61\lambda/(n_{SIL}NA)$. In cryogenic setups, only aspherical lenses with a limited numerical aperture (NA ~ 0.55) have been used [18]. Common microscope objectives which combine many spherical lenses are indeed not suited because of mechanical strain at low temperatures. The Weierstrass SIL geometry [19, 20] effectively compresses the emitted light into a small numerical aperture beam (Fig. 1(a)). More quantitatively, the output emission beam stemming from a point source located at the Weierstrass point of the SIL has a NA of $1/n_{SIL}$. Since the refractive index of the SIL $n_{SIL}$ can be large, a single aspherical lens with relatively low NA is sufficient to collect the half space emission of the emitter with an optical resolution given by $0.61\lambda/n_{SIL}$ [21]. Here we built an objective which is based on a Weierstrass SIL made of ZF14 glass (diameter of 1 mm, $n_{SIL} \sim 1.8$ at 785 nm, matching approximately the average refractive index of the Ac crystal). The flat surface of the SIL coincides with the plane containing the Weierstrass object point (see Fig. 1(a)). The SIL is combined with an aspherical lens with a focal distance of 4.5 mm and a numerical aperture matching $1/n_{SIL}$ (~ 0.55). The overall numerical aperture is close to 1.8 since we are detecting all photons emitted in the SIL half space [21]. By consideration of the losses in optics and filters as well as the collection efficiency of the detector, we estimate that our overall detection efficiency exceeds 10%.

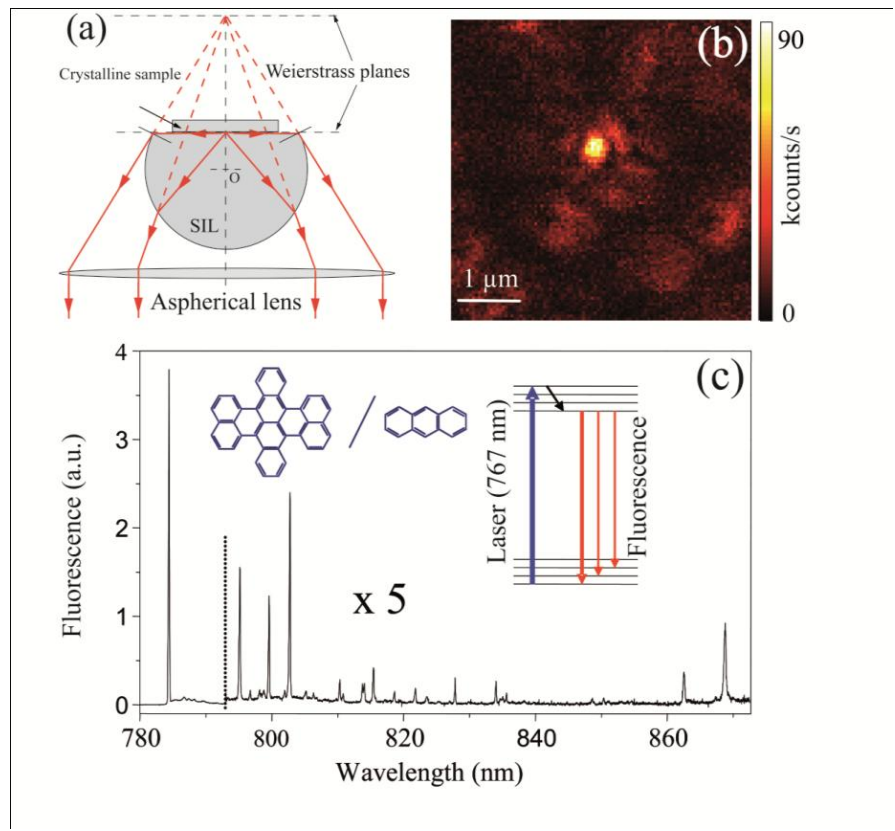

Fig.1. (a) Scheme of our home-built scanning confocal microscope which combines a SIL in the Weierstrass configuration and an aspherical lens. (b) A 5×5 μm² confocal image of single DBT molecules trapped in an Ac crystal at 2 K. The molecules are pumped with a cw laser (at 767 nm) into a vibronic level. This image was recorded with photons emitted on the ZPL, using a combination of steep-edge optical filters having an overall transmission spectral window of a few nanometers width around 785 nm. (c) Emission spectrum of a single DBT molecule, composed of a narrow and intense ZPL, vibrational fluorescence lines, and phonon sidebands. The relative weight of the ZPL in the entire emission spectrum is 33%.

DBT-doped Ac flakes were grown using a co-sublimation technique [22]. They were mounted, together with the SIL, on a combination of a piezo-scanner and slip-stick step positioners, and inserted in a helium cryostat. This combination allows to record confocal images of the entire objective field (estimated to 35×35 μm$^2$ by ray tracing calculations). A Ti:Sapphire laser (767 nm) is used to pump individual DBT molecules into a vibronic state (mode at 300 cm$^{-1}$) [see inset of Fig. 1(c)]. After fast non-radiative relaxation (in the picosecond time scale, as deduced from excitation spectrum linewidths of ~ 40 GHz at 2 K), fluorescence is emitted from the purely electronic excited state, giving rise to a sharp and intense ZPL as well as red-shifted vibrational fluorescence lines. The emitted photons are filtered from the scattered excitation light and sent to single-photon-counting avalanche photodiodes. Single molecules can be imaged with high signal-to-background ratios exceeding ten, even when only the photons arising from the purely electronic ZPL (at 785 nm) are selected, as exemplified in Fig. 1(b). The confocal scanning range is determined using a calibrated pattern located in the Weierstrass plane. The background is mainly due to the emission from out-of-focus molecules. Fitting a single molecule spot with a gaussian profile leads to a point spread function (FWHM) of 300 ± 20 nm for the objective, which is close to the expected diffraction limit at 767 nm (270 nm).

Figure 1(c) shows the emission spectrum at 2 K of a single DBT molecule, recorded with a spectrometer (1800 lines/mm grating, resolution 75 GHz) equipped with a liquid-nitrogen-cooled camera. This spectrum is corrected from the spectral response of the grating and the detector. The intensity ratio between the ZPL and all of its sidebands is governed by the Franck-Condon and Debye-Waller factors. In the emission spectrum displayed in Fig. 1(c), 33% of the single molecule emission intensity originates from the ZPL. This relatively high value makes this system attractive for the efficient delivery of single indistinguishable photons.

Photon correlation measurements require a high detection count rate since the number of coincidences scales as the square of the fluorescence signal for a given acquisition time. In order to compare the collection efficiency of our home-built microscope with a commercial microscope objective (NA = 0.95), we performed optical saturation studies in both configurations, selecting the ZPL photons. Figure 2(a) shows a plot of the fluorescence signal $S$ of this molecule as a function of the excitation power $P$. The detected count rate of the ZPL photons is in agreement with the saturation law:

$$S/S_\infty = \frac{P/P_{sat}}{1+P/P_{sat}},$$

where $S_\infty$ is the saturated count rate and $P_{sat}$ the saturation power. For this molecule we find $S_\infty$ ~ 180 kcounts/s, which is more than four times larger than that obtained with the air microscope objective (~ 40 kcounts/s).

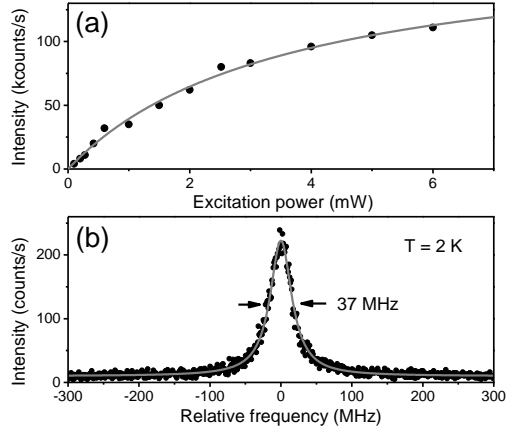

Fig. 2. (a) Saturation plot of the ZPL fluorescence signal as a function of the excitation power. A fit of the data with the saturation law gives a saturated count rate of 180 kcounts/s and a saturation power of 3.5 mW. (b) ZPL excitation spectrum of a single DBT molecule at 2 K, at an excitation intensity well below the saturation intensity. A fit with a Lorentzian profile gives a homogeneous linewidth of 37 MHz.

We also measured the fluorescence excitation spectrum of the purely electronic transition at low temperature. The frequency of a cw monomode laser (spectral resolution 1 MHz) is swept over hundreds of MHz around the ZPL, while collecting red-shifted fluorescence photons ($\lambda$>790 nm) through a long-pass filter. Figure 2(b) shows the ZPL excitation spectrum of a single DBT molecule at T = 2 K in the low saturation regime. It displays a Lorentzian profile with a linewidth (FWHM) of 37 MHz, which shows that the optical coherence decay rate reaches its lower bound determined by the excited state lifetime. Indeed, fluorescence decay measurements lead to a lifetime of 4.5 ns.

We studied the fluorescence intensity correlation function measured with ZPL photons emitted by a single DBT molecule, using a Hanbury-Brown and Twiss setup. We first recorded the start-stop coincidence histogram $C^{(2)}(\tau)$ with a single molecule under a cw excitation to the vibronic state. As shown in Fig. 3(a), a strong photon antibunching is observed, which demonstrates single photon emission. The experimental histogram can be fitted with an exponential rise:

$$C^{(2)}(\tau) = C^{(2)}(\infty)\left\{1 - b\exp\left[-\frac{|\tau|}{\tau_f}\left(1+\frac{P}{P_{sat}}\right)\right]\right\},$$

where $P$ is the excitation power, $\tau_f$ and $P_{sat}$ are respectively the excited-state lifetime and its saturation power. Note that at low temperatures, this expression is valid only for excitation rates much lower than the vibronic level relaxation rate. The triplet manifold is neglected here since the intersystem crossing yield is smaller than $10^{-7}$ [23]. Figure 3(a) exemplifies a coincidence histogram recorded at $P/P_{sat} = 0.24$. From this histogram we deduce $\tau_f = 4.5$ ns and a dip amplitude b of 82%. Zero delay coincidences are consistent with the background level.

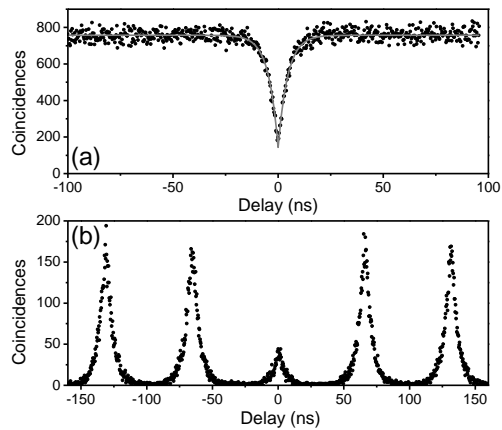

Fig.3. Histograms of time delays in a start-stop experiment performed with ZPL photons. (a): CW excitation (power 1 mW), over an integration time of 600 s. A fit of the experimental correlation curve gives a relative dip of 0.82 and an excited-state lifetime $\tau_f = 4.5 \pm 0.1$ ns. (b): Pulsed excitation (average power 3.5 mW), over an integration time of 1600 s. The ratio between the areas of the central peak and one of the lateral peaks is 22 %. The exponential decay of the lateral peaks gives the excited-state lifetime $\tau_f = 4.7 \pm 0.3$ ns.

We then pulsed the excitation to trigger the emission of single-photons from an individual molecule, using a mode-locked Ti-Sapphire laser (300 fs pulse duration, 16 MHz repetition rate). As shown in Fig. 3(b), the coincidence histogram has the expected peak pattern, given the periodic emission of photons by the molecule. For a perfect single photon emission the central peak should vanish. In our case, the relative weight of the central peak is weak (22 %) as compared to one of the lateral peaks. The residual zero delay coincidences are attributed to events involving the background [4]. Because the pulse duration is 4 orders of magnitude shorter than the emitting state lifetime, the probability to generate two photons per pulse is negligible (less than 1 %). From the exponential decay of the lateral peaks we can deduce the excited-state lifetime. We found $\tau_f = 4.7 \pm 0.3$ ns, a value consistent with that deduced from the ZPL excitation linewidths.

In summary, we built a solid immersion confocal objective which images single molecules with an optical resolution of 300 nm and efficiently collects their emission photons at liquid helium temperatures. A high repetition rate source of near-infra-red single photons is realized. Since the photons are selected from the strong zero-phonon line of DBT molecules at 2K, they should be indistinguishable. A coalescence experiment is the further step.

This research was funded by the Région Aquitaine and the Agence Nationale pour la Recherche.